\documentclass{iau} 
\usepackage{graphicx}

\title[Chemo-dynamical evolution of the inner Milky Way] %% give here short title %%
{Using N-body simulations to understand the chemo-dynamical evolution of the inner Milky
Way}

\author[E. Athanassoula]   %% give here short author list %%
{E. Athanassoula} 
\affiliation{Aix Marseille Univ, CNRS, LAM, Laboratoire d'Astrophysique de Marseille, Marseille, France \\ email: {\tt lia@lam.fr}} 

\pubyear{2017}
\volume{IAU334}  %% insert here IAU Symposium No.
\jname{Rediscovering our Galaxy}
\editors{C. Chiappini, I. Minchev, E. Starkenburg, M. Valentini, eds.}
\begin{document}

\maketitle

\begin{abstract}  I present examples of how chemo-dynamical N-body
  simulations can help
  understanding the structure and evolution of the inner
  Galaxy. Such simulations reproduce the observed links
  between kinematics, morphology and chemistry in the bar/bulge region
  and explain them by the self-consistent cohabitation of a number of
  components. 
  Galactic archaeology, applied to simulation snapshots, explains the
  sequence in which the stars of the various components were formed.
  The thick disc stars form earlier than those of the thin disc and in
  a much shorter time scale. The bar in 
  the thick disc is horizontally thicker than that of the thin disc
  and has a different vertical morphology. The Galaxy's   inner 
  disc scalelength is much smaller than what is expected from 
  nearby galaxies of similar stellar mass.

\end{abstract}

\firstsection % if your document starts with a section,
\section{Introduction}

The unprecedented quality and quantity of the data on the Milky Way
(MW), from Gaia and from the accompanying ground-based surveys (BRAVA, RAVE, SEGUE, ARGOS, APOGEE, LAMOST, Gaia-ESO, VVV-ESO, GALAH, WEAVE, 4MOST, etc.) has instigated a large number of N-body simulations, which include gas, star formation, feedback and cooling, and, in a few cases, chemical evolution. It is not possible to discuss them all in this review, so I focus on some specific recent results, %which, to my eyes, are particularly important, or promising, 
thereby forsaking any attempt for completeness. %Furthermore, I have focused on the bar/bulge region and, albeit only concerning the surface brightness radial profiles,  to the inner disc. 
Furthermore, I do not discuss here any cosmological simulations, since they are the subject of Scannapieco's review (this volume),
and I barely touch on results on the solar neighbourhood, as they are extensively 
discussed e.g. in the reviews by \cite{Minchev.17} and by Martig (this volume). 

For all the figures and analyses made
specifically for this review I used the survey of several hundred high resolution 
simulations initiated with the work described in \cite{Athanassoula.RPL.16} (hereafter A16; see also
\cite{Rodionov.AP.17} and \cite{Athanassoula.RP.17}, hereafter A17)
and to which I will refer here as the ARPL simulations.      
Each of these starts off with two idealised but cosmologically
motivated protogalaxies, composed initially only of dark matter and hot
gas, and set on an orbit which will bring them sooner or later to
merge. Their mass ratio is such as to lead to a major, or intermediate
merger. Stars are not present in these initial conditions, but start
forming well before the merging, creating a protodisc in each
protogalaxy. The merging destroys these two discs, while the two dark
matter and the two gaseous haloes merge. A new disc is then formed
gradually -- inside out horizontally and outside in vertically -- from accreting halo
gas. By including in the formation scenario such a merger (which, in current cosmological models, is 
expected to be part of the formation history of a considerable fraction of MW-like galaxies), 
%occur in a very considerable fraction of cases  at relatively large lookback times%(\cite{Stewart.BWMZ.08}, and references therein)
these simulations become more realistic, and can naturally
account for components such as a classical bulge or a thick disc with realistic properties and chemistry distinct from that of the thin disc. It also explains better the observations, as e.g. the kinematics of low metallicity stars (\cite{Ness.P.13}) and A17), the formation of the various types of radial density profiles etc. The large number of ARPL simulations available also allows a more homogeneous approach to many important issues on the structure and  evolution of MW-type galaxies. 

Last but not least, I discuss some specific comparisons with external galaxies, because it should not be forgotten that our Galaxy is one of many, and thus could not have properties at odds with those of other disc galaxies of similar type and mass. 

\section{Stellar density distribution}
\label{sec:dscalelengths}

\begin{figure}[b]
% \vspace*{-2.0 cm}
\begin{center}
 \includegraphics[width=5.4in]{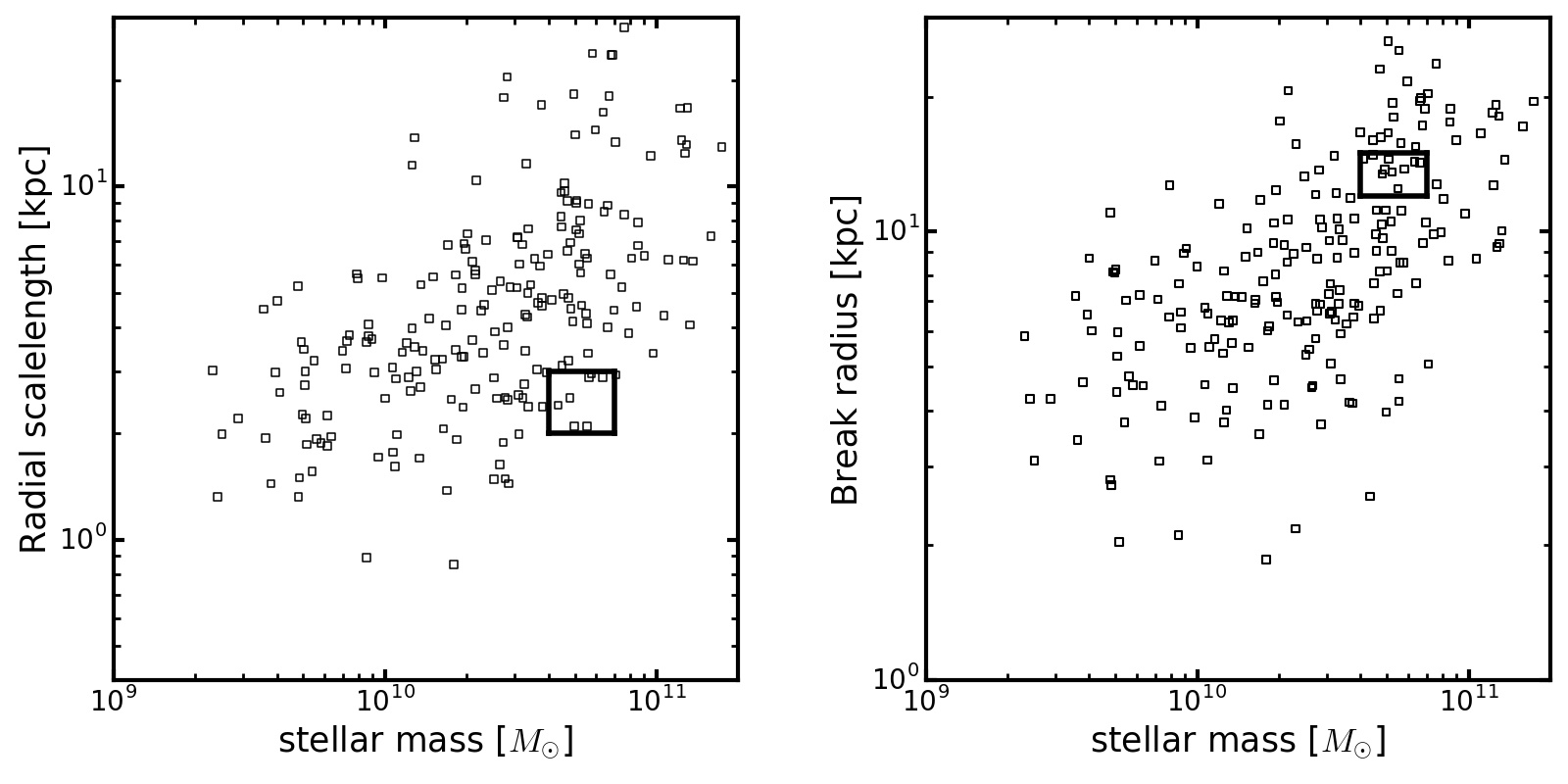} 
%\includegraphics[width=5.4in]{fig_scaling_mass3_color.jpg} 
% \vspace*{-1.0 cm}
 \caption{Left: %Comparison of the inner 
 Inner disc scalelength of galaxies from the S$^4$G and NIRS0S samples as a function of their stellar mass. %(open symbols)  to those of  APRL
%   simulations with MW-like stellar masses (filled symbols). 
The open rectangle outlines the region where the values of the MW are
   expected to lie.  Right: Similar plot, but now for the break radius.}
   \label{MW-inner-disc-scalelength}
\end{center}
\end{figure}

%There are three types of disc surface density profiles, namely Type I, II
%and III. In the first type the exponential profile of the main part of
%the disc extends all the way to the outermost region. On the contrary in
%types II and III there is a break, beyond which there is a second
%exponential profile. This describes the outer disc
%and can have a slope which is smaller (for types II) or larger (for
%types III) than that of the inner disc. Such profiles have been found
%in a very large number of both simulations and observations.   

Observations have shown that the vast majority of the stars in MW-type spiral galaxies lie in a disc component which has an exponential radial density profile with, or in a few cases without, a break separating the inner from the outer disc. Our Galaxy is no exception. These structures, as well as their properties as a function of lookback time and stellar age, have been well reproduced by simulations (e.g. by %????, \cite{Roskar.DSQ.08}, 
\cite{Radburn-Smith.RDD.12}, \cite{Aumer.White.13},
\cite{Martig.MF.14}, A16, \cite{Aumer.Binney.Schonrich.16}, references
therein, and as is shown in more detail in ongoing work in
collaboration with Peschken et al.).  

The left panel of Fig.~\ref{MW-inner-disc-scalelength} shows the inner
disc scalelengths, $R_{in}$, %(i.e. the scalelength of the part within
                             %the break) 
of the galaxies in the \cite{Laine.LS.14} sample from the  S$^4$G (\cite{Sheth.S4G.10}, \cite{Salo.S4G.15}) and NIRS0S (\cite{Laurikainen.SBK.11}) samples, as a function of their stellar mass ($M_*$). 
%to the measurements of the same quantity for a sample of MW-like  ARPL simulations. As the ultimate goal here is
%the comparison with the MW, I only use MW-like simulation with . Together with the left and middle
%panels (for the outer and break radii), this shows that the simulation
%sample has scalelengths and break radii which cover the same range
%values as MW-type galaxies.
The rectangle outlines a region in which the MW values should be located, i.e. with an $M_*$ of 4 -- 7 $\times$ 10$^{10}$  M$_\odot$  
and an $R_{in}$ of 2 -- 3 kpc. Note that for the sample in Fig.~\ref{MW-inner-disc-scalelength}, only 7 out of 49 external 
galaxies  with a MW-like stellar mass (or 4 out of 30, if the mass
limit is 5 -- 7 $\times$ 10$^{10}$  M$_\odot$), have the required
$R_{in}$, all the remainder having considerably larger values, i.e. our Galaxy has a short $R_{in}$ for its
mass. %Its value is not outside the range outlined by theS$^4$G, it is simply much smaller than the average. Indeed only . 

The MW break radius is of course less accurately known than the inner
disc scalelength, but still, the right panel of Fig.~\ref{MW-inner-disc-scalelength} shows clearly that it is much nearer to that of the average external galaxy. Thus the MW inner scalelength is much smaller than expected both for its $M_*$ and its radial extent, the latter measured by the break radius. %Before reaching the conclusion that our Galaxy is peculiar, one should, however, examine the various ways the disc scalelength is measured. 
  
%In the oral presentation I also discussed the scalelengths and break
%radii as a function of lookback time and of the age of the stellar
%population in which this quantity is measured, comparing with other
%simulations (e.g. \cite{Aumer.White.13}, \cite{Martig.MF.14}, A16, \cite{Aumer.Binney.Schonrich.16}, ????REFS????) and observations. 
Swing amplification (\cite{Toomre.81}) -- as applied to external galaxies (\cite{Athanassoula.BP.87}) and using the parameter values given by \cite{Bland-Hawthorn.Gerhard.16} -- predicts that at the solar region the spiral arm multiplicity ($m$) is around 4 - 5. This is in agreement  with the observations of  \cite{Georgelin.Georgelin.76}, and does not %necessarily 
disagree with an $m$=2 in the inner parts. Although a number %of examples 
of such disc galaxies are known
(e.g. \cite{deVaucouleurs.Pence.78}, \cite{Efremov.11} and references
therein), they are still a minority, compared to other barred galaxy
morphologies. This again implies that the MW is similar to some
external galaxies, but still far from the average barred galaxy. Note
that NGC 5012 -- which is the only one external galaxy from the sample
in Fig.~\ref{MW-inner-disc-scalelength} with an $M_*$, an $R_{in}$ and
a break radius compatible with those of the MW -- has also 4 - 5 spiral arms outside the central region.

\section{Galactic archaeology}
\label{sec:garchaelogy}

Galactic archaeology uses properties (such as kinematics, spatial
distribution and surface chemical abundances) of present day stars, to understand
the formation and evolution of our Galaxy. The
stars can be divided in groups according to their age (or its proxy)
and the differences between the properties of the various groups 
is used to outline the Galaxy's formation and evolution. 

%ir It was introduced by
%\citeauthor{Freeman-BH-02} and extended and extensively used since
%(see e.g. ??? in this volume). The task is considerably complicated by
%the fact that stars do not necessarily stay at or near their birth
%position, but may migrate by a considerable distance. It is thus
%useful to apply Galactic archaeology techniques to mock (simulated)
%galaxies, where also direct knowledge on the formation and
%evolution is available. 

To test Galactic archaeology techniques, A16 applied them to the final snapshot of one of their simulations and
divided the stars into groups as a function of their age. %Contrary to some previous works, t
The dividing times were chosen to be
landmark times of the evolution. The two main
such times are the beginning and the end of the merging
period, i.e. of the time span during which the two
protogalaxies merge. Although these two times can not be precisely calculated,
all reasonable estimates will give the same qualitative
results and very similar quantitative ones. The end of the merging period can be
considered the same as the beginning of the disc formation ($t_{bd}$). 
A16 studied five age groups separately using face-on and end-on morphology,
radial and vertical surface density profiles and the circularity of the individual
stellar orbits. The latter is defined as the $z$ component of the angular momentum, normalised by the angular
momentum of a circular orbit with the same energy (\cite{Aumer.White.13}, A16). 

Before the merging period starts, the two protogalaxies are apart, 
forming their stars in isolation. These stars undergo violent
relaxation during the merging and, viewed globally at the end of the simulation,
exhibit properties similar to those of a classical bulge or stellar halo. Thus
gas-rich major or intermediate mergers can explain the formation of these components,
whose stars have formed during the initial stages of the evolution. 

After $t_{bd}$, stars form gradually from accreting halo gas and within a thin disc component. Thus the thin disc is  formed during the later secular evolution times.

The evolution of stars born in the intermediate times, i.e. during the merging period, is more
complex. Most stars forming during the beginning
of that period contribute mainly to the
classical bulge and stellar halo, but most stars born in the later stages of the merging period contribute to the thick disc. There
is no real discontinuity at any time, but a continuous shift from
spheroids to a thick disc component and then to a thin disc. 

Simulations such as in A16 show how the various stages of formation and evolution
interconnect, shifting from merger-driven to secularly-driven regimes. They also
argue that the formation is generally from inside out horizontally,
and from outside in vertically. Thus, the thick disc forms first and in a
relatively short time scale of the order of a Gyr, while the thin
disc starts forming later and continues its growth for a much more
extended period of time, of the order of 10 Gyr. All this is in agreement with the observed bimodality in e.g. the [$\alpha$/Fe] - [Fe/H] relation, and could provide an explanation of the existence of two distinct trends. Thus this galaxy formation picture leads naturally to the star formation history necessary for the chemical evolution model of \cite{Chappini.MG.97}.

Note also that some of the stars born in the thin disc sufficiently early on  will be heated by encounters with e.g. spiral arms,
giant molecular clouds, or globular clusters and thus can become members of the thick disc, as will be discussed in more detail elsewhere. 
     
\section{Bars}

\subsection{General}

Bars drive considerable secular evolution in disc galaxies, via angular momentum and mass redistribution. The former takes place via exchanges between the various resonances in the disc and halo (\cite{Athanassoula.02}, %\cite{Athanassoula.03}, 
or for a general review \cite{Athanassoula.13} and references therein).  As most disc galaxies, the MW has a bar. Its length, however, is shorter than what would be expected for galaxies with a MW-like $M_*$. This is not necessarily a new particularity of the MW, but may simply be due to the fact that the $R_{in}$ of the MW is small, as discussed in Sect.~\ref{sec:dscalelengths}. Thus the bar length, as measured in kpc, would indeed be smaller than expected, but, when measured in inner disc scalelengths, would be much nearer to expectations.      
This will also be the case for the resonant radii, so that the radial extent of the region which is strongly influenced by the bar would be similar to that of other barred galaxies, when measured in disc scalelengths. 

%?? It should also be mentioned that this radial extent could be increased if the 

%\subsection{Interaction between the bar and the halo}

\subsection{A bar in the thick disc}

\begin{figure}[b]
% \vspace*{-2.0 cm}
\begin{center}
 \includegraphics[width=5.4in]{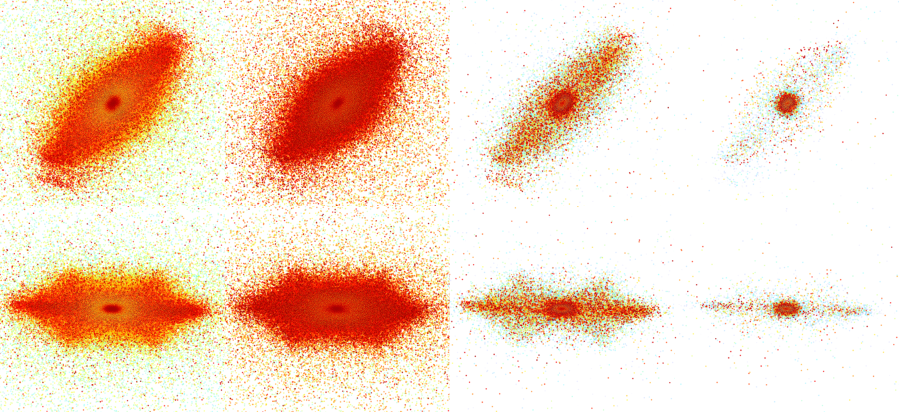} 
% \vspace*{-1.0 cm}
 \caption{Comparison of the bar morphology for stars in different age ranges (see text). The upper panels in each column give the face-on view with the bar roughly on the diagonal, and the bottom ones  the side-on view. The leftmost column of panels includes bar stars of all stellar ages. For the remaining three columns of panels, the stellar age decreases from left to right. The second column from the left includes mainly the stars in the thick disc and the third and fourth columns mainly the stars in the thin disc. Most of them are in the third panel, except for the youngest (not older than one Gyr) which are in the rightmost panel. Note the strong difference in morphology between the bars of different age stars.}
   \label{thin-thick-bar}
\end{center}
\end{figure}

Nearly 35 years ago, using N-body simulations with the resolution allowed by the computers at the time, \cite{Athanassoula.83} predicted that the bar shape depends on the stellar velocity dispersion; cold discs forming thinner bars than hot ones. She also showed that if there are two stellar populations, of different stellar velocity dispersions, the cold one will create a thin bar, while the hot one will create a thicker bar with the same length and pattern speed. These predictions have been confirmed by a number  of more recent and technically superior simulations (\cite{Athanassoula.03}, \cite{Bekki.Tsujimoto.11}, A16, \cite{Debattista.NGFZM.17}, \cite{Fragkoudi.DMHGCKS.17}). It is thus expected that, in the MW, both the thick and the thin disc would grow a bar. Whether these can be considered as two separate bars, dynamically locked to each other, or one single bar where stars of different ages outline different morphologies, depends on whether one believes there is a clear separation between the two discs, or a more or less continuous transition.    

Most of the above works have very idealised initial conditions, in
which the thin and the thick disc populations are present and
separated already in the initial conditions of the simulations. It is
then trivial to distinguish between the two bars and to find to which
bar a given particle belongs. This, however, is not the case in
observations, where  the two populations have to be distinguished
e.g. by their age, or their metallicity. Simulations with less
idealised initial conditions, where the thin and the thick discs do
not pre-exist in the initial conditions, but form during the
evolution, are closer to observations. This is the case with the ARPL
simulations. These present a number of advantages. More specifically,
gas is accreted from the galactic halo and forms stars, so that the
age of each star is trivially calculated all through the simulation
and there is no need of proxies. Moreover, the thick disc forms
naturally in this scenario and has properties which agree with
observations (A16). In fact, the third and fourth panels (from the
left) of  Fig. 4 in A16 have already shown that the thick disc
population, viewed face-on at the end of the simulation, has a central
thick bar of oval shape, which viewed side-on (i.e. edge-on, with the
line of sight along the bar minor axis) has a boxy shape. On the contrary, the younger stars, born in the thin disc after those in the thick disc, show face-on a much thinner bar and side-on an X shape. For further discussion on this see Sect. 3.9 of A16.

The example discussed in A16 has a short and relatively weak bar. I
will extend their analysis here using a simulation and snapshot
with a longer and stronger bar, which increases the S/N ratio
and illustrates better the  effects I wish to show. I follow the same
procedure as in A16, %, splitting the stars into bins according to their age and using similarly motivated time boundaries between the groups. 
but contrary to it I include in the decomposition only stars which
at the end of the simulation are part of the bar, as found according
to their location and kinematics. The result is shown in
Fig.~\ref{thin-thick-bar}, while a colour version of this plot can be
seen in slide 20 of my oral presentation
(https://iaus334.aip.de/pdf-of-the-talks/). In the latter the colour
represents age, the oldest in each group in red and the youngest in
blue. It is clear that the oldest bar population -- which contains the
stars formed in the later part of the merging -- are distributed
face-on within a thick oval shape, and edge-on extend to large
distances from the equatorial plane, outlining a boxy-like shape. On
the contrary, the younger ones, born during the thin disc formation
time, are distributed in a much thinner bar shape face-on and more
X-like shape edge-on. This is in good agreement with what is expected from previous work (see in particular A16).

The youngest stars, of age less than a Gyr, form the thinnest of
all distributions, both horizontally and vertically. It is quite possible that these stars may form a vertically 
super-thin bar as the one discussed by \cite{Wegg.GP.15}. A full appraisal of
this suggestion, however, will necessitate a baryonic softening of the
order of at most 10 pc and better around 5 pc, while most of the ARPL
simulations have a softening of 25 pc, which, although considerably smaller
than most other simulations addressing this question, is still 
larger than what is required. %????Note also that the required low softening value introduces a number of other technical difficulties so will not be discussed further here.  

Hence these simulations show that the thick disc bar morphology is different from that of the thin
disc, resembling, face-on, more a fat oval. They also do not have ansae and the
thick disc in general shows no spirals, unless of very low amplitude
(A16). These predictions should be testable for our Galaxy using data from Gaia and the accompanying ground-based surveys.

\section{Coupling chemical evolution with kinematics and morphology in the bar/bulge region}

\begin{figure}[b]
% \vspace*{-2.0 cm}
\begin{center}
 \includegraphics[width=5.4in]{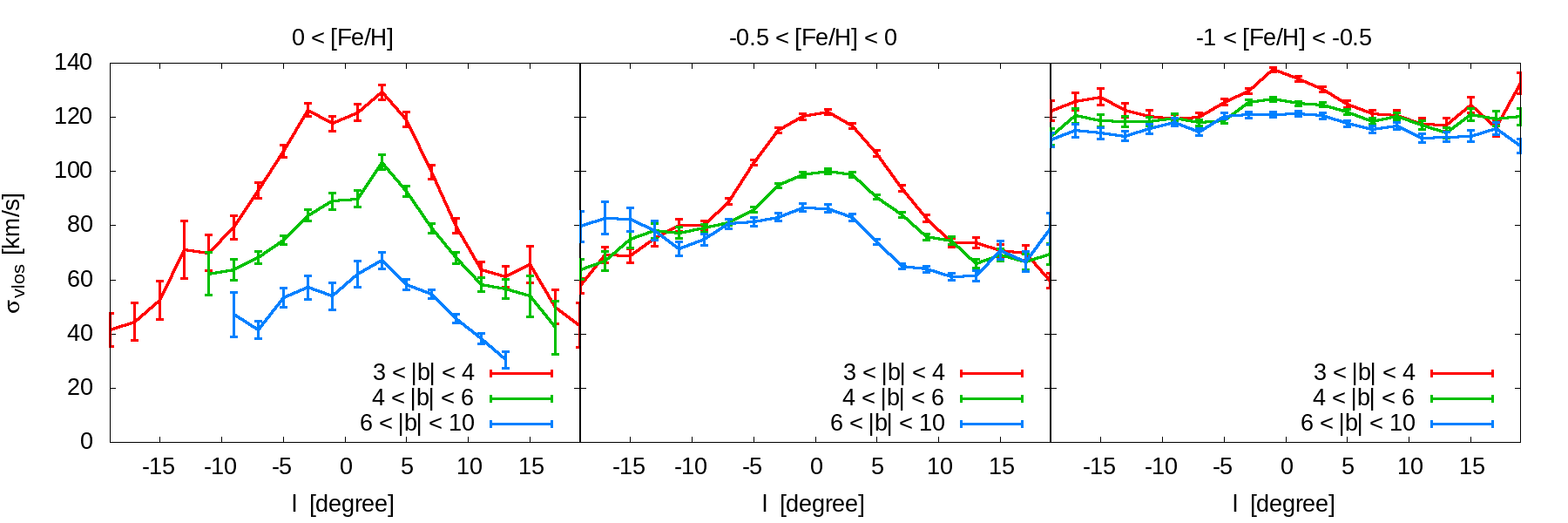} 
% \vspace*{-1.0 cm}
 \caption{Link between the metallicity and kinematics: Line-of-sight
   velocity dispersion as a function of longitude. Stars are divided
   into groups as a function of their latitude and of their
   metallicity (HM to the left, LM stars to the right, and IM in the
   middle panel). Reproduced from Fig. 2  of A17.}% + take out middle panel AB???}
   \label{Melissa-fig6-sims}
\end{center}
\end{figure}

The high quality data % that has been% and will be gradually 
delivered by the spectroscopic surveys accompanying Gaia brought %(or will bring) 
a lot of information on the element abundances of MW stars.  This can be used to set further constraints on the models and in particular on the new N-body simulations. 

A number of the early chemo-dynamical simulations included no gas
component and thus no information on stellar ages and no possibility
of modelling the chemical evolution. The metallicities and abundances
were replaced by proxies. This can be hazardous, but still such codes
have been often used and allowed some interesting initial
insights. With the advent of N-body codes with a full hydrodynamic
treatment, i.e. including star formation, feedback and cooling, a few
codes made the extra steps necessary to include chemical
evolution. This brought much tighter constraints on the models,
opening up a whole new realm of possibilities. %A large number of N-body simulations studied the solar neighbourhood, see e.g. the reviews by \cite{Minchev.17} and by Martig (this volume), and references therein. 
I will focus here only on the bar/bulge region.  

Observations clearly showed that the metallicity, kinematics and spatial distribution of the stars are strongly coupled to each other. \cite{Babusiaux.16}, using data from a number of sources, %, including her own, 
plotted the line of sight velocity dispersion ($\sigma$) as a function of the absolute value of the latitude separately for low metallicity (LM, -1.0$<$[Fe/H]$<$-0.5) and high metallicity (HM, 0.$<$[Fe/H]$<$0.5) stars, neglecting the intermediate ones (IM), 
and finds they have a quite different behaviour.  For the LM stars, $\sigma$ is roughly constant, showing no dependence on latitude. On the contrary, for the HM stars $\sigma$ shows a strong decrease with increasing latitude reaching much smaller values of $\sigma$ than the LM at high latitudes.  Using one of the very high resolution ARPL simulations, A17 reproduced and discussed this behaviour (see left panel of Fig. 1 in A17).          

In similar, but more in depth and more extended studies,
\cite{Ness.P.13}, their Fig. 6) using the ARGOS sample
(\cite{Freeman.P.13}) and \cite{Zasowski.NGMJM.16}, their Fig. 9)
using APOGEE data (\cite{Majewski.P.17}) showed the link between the
metallicity and the kinematics, not only as a function of latitude,
but also as a function of longitude. At low latitudes the $\sigma$ of
HM stars shows a clear, strong maximum at zero longitude, declining
sharply with increasing absolute value of longitude. Still for HM
stars, but now at larger distances from the Galactic equatorial plane,
$\sigma$  shows lower values and a much shallower decrease with absolute longitude%, or hardly any decrease at all
. The LM stellar population has higher values of $\sigma$ and much
less of a decrease than the HM one. Populations of intermediate
metallicity have also intermediate kinematics. These qualitative
behaviours are found in both the ARGOS and the APOGEE data. With their very high resolution simulations, A17 reproduce all these trends and explain them by linking both the metallicity and the kinematics to the  stellar origins. Thus, stars born before or during the merging time will have low metallicities and high velocity dispersions. As mentioned in Sect.~\ref{sec:garchaelogy}, they constitute the stellar halo, the classical bulge (whenever present) and the thick disc. On the other hand, stars born during the secular evolution time will form mainly in the thin disc and will have higher metallicities and lower $\sigma$ values.  Thus there is a link not only between kinematics and metallicity, but also between these quantities and morphology. Indeed, metallicity-defined populations are found to have strikingly different spatial distributions. The LM stars are distributed in a spheroidal-like shape highly flattened in the vertical direction, and including a thick, low-density disc. They do not
show any, or very little, X-shaped structure, while the HM stars outline only the X-shape and a clear underlying disc component (Figs. 3 and 4 of A17), all in good agreement with observations.
%.in good agreement with observations \cite[e.g.][]{Ness.P.12, Uttenthaller.SNRLC.15, Rojas-Arriagada.P.14, Kunder.P.16}. 

Note that the simulations reproduce very well {\it qualitatively} all
the observed trends. It is not useful or even appropriate to make
quantitative comparisons, as non-negligible quantitative differences
exist even between the afore mentioned observational studies,
presumably due to the differences in the selection criteria of the two
samples. Also such differences can be due to small differences in the
chemical models (A17). The most important reason, however, is simply
that we do not yet have a simulation which gives a sufficiently good
match to al the MW structural and dynamical properties.

\section{A few concluding words}

I briefly described a few examples demonstrating how useful high
quality N-body simulations  can be in explaining the formation,
structure and evolution of MW-like galaxies. Such simulations should
be of high resolution, and include gas, star formation, feedback and
cooling, and preferably be coupled to a chemical evolution code, thus
giving results on metallicity and element abundances and allowing
comparisons with the results of the numerous ground-based surveys
accompanying Gaia. It will thus be possible to set very tight
constraints on the models, particularly for the formation and
evolution of the various stellar populations and components. The
examples discussed here, and/or in the oral version
(https://iaus334.aip.de/pdf-of-the-talks/),  include galactic
archaeology examples leading to a thick disc formation scenario, the
formation and evolution of bars, the distinction between bars forming
in the thin and the thick disc, the role of the halo in the evolution,
the formation and evolution of exponential radial density profiles
with and without breaks,  and, last but not least, the links between
the kinematics, metallicity and morphology of the various stellar
populations in the bar/bulge region. Obviously this list is far from
exhaustive, but is only meant to give a flavour of the available
possibilities. %as N-bodies have recently reached the necessary
               %resolution and become sufficiently realistic to permit
               %comparisons with MW observations from GAIA and the accompanying ground based surveys. ????

It is also important to keep in mind that our Galaxy is just one disc galaxy among many others. It is unique only in the sense that, as we are in it, we can observe individual stars at a level which can not yet be achieved for most external galaxies. However, it should not be unique, either regarding its formation and evolution, or regarding its structure and contents. Note that numerous external galaxies have been well observed, thus providing invaluable statistical knowledge, while being exempt of the geometrical problems which we have in the MW due to our location in it. Thus a lot has to be gained by close collaborations between people working on our Galaxy and those working on external galaxies.

%The most appropriate concluding sentence is that we are very lucky to work in the pursuit of understanding the %formation and evolution of MW-type galaxies  during these very exciting times!     

\acknowledgements{I thank the organisers for inviting me to 
%review the chemo-dynamical evolution of disc galaxies in
this interesting and inspiring meeting. I thank the CNES for financial support and my collaborators A. Bosma,  
J.C. Lambert, N. Peschken, N. Prantzos, and S. Rodionov, for many useful
discussions and/or help with technical aspects. I acknowledge use of 
HPC resources from GENCI/TGCC/CINES and from Mesocentre of Aix-Marseille-Universit\'e. 

%This work was granted access to the HPC resources of
%[TGCC/CINES/IDRIS] under the allocation 2016047665 ????and??? and by  
%It was also granted access to the HPC resources of Aix-Marseille
%Universit\'e financed by the project Equip@Meso (ANR-10-EQPX-29-01) of
%the program ``Investissements d'Avenir'' supervised by the Agence Nationale de la Recherche.}

\end{document}